\documentclass[apj]{emulateapj}
\pdfoutput=1

\newcommand{\appropto}{\mathrel{\vcenter{
  \offinterlineskip\halign{\hfil$##$\cr
    \propto\cr\noalign{\kern2pt}\sim\cr\noalign{\kern-2pt}}}}}

\usepackage{pslatex}
\usepackage{graphicx}
\usepackage[flushleft]{threeparttable}
\shorttitle{SZ signal from Quasar Hosts}
\shortauthors{Dutta Chowdhury \& Chatterjee}
\journalinfo{Accepted into \apj, March 2, 2017}

\begin{document}
\title{Sunyaev-Zel'dovich Signal from Quasar Hosts: Implications for Detection of Quasar Feedback} 
\author{Dhruba Dutta Chowdhury$^{1,2}$, Suchetana Chatterjee$^{1}$}
\affiliation{$^{1}${Department of Physics, Presidency University, Kolkata, 700073, India},\\ $^{2}${Department of Astronomy, Yale University, New Haven, CT 06511.}}

\email{dhruba.duttachowdhury@yale.edu}
\email{suchetana.physics@presiuniv.ac.in}

\begin{abstract}
Several analytic and numerical studies have indicated that the interstellar medium of a quasar host galaxy heated by feedback can contribute to a substantial secondary signal in the cosmic microwave background (CMB) through the thermal Sunyaev-Zel'dovich (SZ) effect. Recently, many groups have tried to detect this signal by cross-correlating CMB maps with quasar catalogs. Using a self-similar model for the gas in the intra-cluster medium and a realistic halo occupation distribution (HOD) prescription for quasars we estimate the level of SZ signal from gravitational heating of quasar hosts. The bias in the host halo signal estimation due to unconstrained high mass HOD tail and yet unknown redshift dependence of the quasar HOD restricts us from drawing any robust conclusions at low redshift ($z<1.5$) from our analysis. However, at higher redshifts ($z>2.5$), we find an excess signal in recent observations than what is predicted from our model. The excess signal could be potentially generated from additional heating due to quasar feedback.
\end{abstract}

\section{Introduction}

The measurement of the temperature fluctuations in the cosmic microwave background (CMB) maps at large and small angular scales have proven to be one of the most powerful probes of cosmology \citep[e.g.,][]{spergeletal03, spergeletal07, reichardtetal09, dunkleyetal11, dasetal11a, dasetal11b, sehgaletal11, sherwinetal11, hlozeketal12, reichardtetal12, sieversetal13, calabreseetal13, storyetal13, planckcollaboration13XVI, planckcollaboration13XX, planckcollaborationetal13XXII, planckcollaboration13XXIV, dasetal14, houetal14, planckcollaboration15XIII, planckcollaboration15XIV,  planckcollaboration15XVII,planckcollaboration15XVIII, georgeetal15}. 

The detection of the secondary fluctuations from astrophysical sources via cross-correlation techniques with other wavebands have been the primary focus of many microwave experiments \citep[e.g.,][]{diegoetal03,hirataetal04,afshordietal04,chengetal04,padmanabhanetal05,hoetal08,chatterjeeetal10, fengetal12,kovacsetal13, giannantonioetal14, munshietal14, cabassetal15, bianchinietal15, ferraroetal15, ruanetal15, grecoetal15, crichtonetal16, spaceketal16, z&m16, verdieretal16, soergeletal16}. The Sunyaev-Zeldovich effect \citep [SZ;][]{s&z70,s&z72}, arising from inverse Compton scattering of CMB photons happens to be the dominant secondary signal at angular scales of an arcminute. The SZ effect creates a spectral distortion in the CMB and serves as a probe for accumulations of hot gas in the Universe (see \citealt{carlstrometal02} and references therein). 

The major SZ signal comes from galaxy clusters which are reservoirs of keV energy electrons \citep[e.g.,][]{birkinshawetal78,joyetal01,hincksetal10,staniszewskietal09,planckcollaborationetal15XL}. However, apart from the cluster signal several small scale astrophysical sources produce measurable SZ distortions in the CMB \citep[e.g.,][]{mcquinnetal05, ilievetal07, whiteetal02, majumdaretal01, aghanimetal00, dezottietal04, rosagonzalezetal04, massardietal08, b&l07, ohetal03, aghanimetal08, reichardtetal12, archidiaconoetal12,vandevoortetal16}. The SZ signal from these sources provides a new observational tool to study the role of baryonic physics on structure formation. 
 
A generic astrophysical signal that is related to the SZ distortion from hot gas surrounding an active galactic nucleus (AGN) has been studied by several authors using analytic and numerical tools \citep[e.g.,][] {n&s99, aghanimetal99, yamadaetal99, lapietal03, plataniaetal02, roychowdhuryetal05, c&k07, scannapiecoetal08, chatterjeeetal08, zannietal05, sijackietal07}. The first attempt to detect this signal was carried out by \citet{chatterjeeetal10} who used a technique of stacking quasars from the Sloan Digital Sky Survey \citep[SDSS;][]{schneideretal10} onto the CMB temperature maps from the Wilkinson Microwave Anisotropy Probe \citep[WMAP;][]{bennettetal03}. 

\citet{chatterjeeetal10} reported a tentative detection of the signal within the noise sensitivity and the angular resolution of WMAP. Recently, several other teams have tried to detect this signal using the Planck Surveyor Satellite \citep[R15 hereafter][V16 hereafter]{ruanetal15, verdieretal16}, the Atacama Cosmology Telescope (ACT; \citealt{crichtonetal16}, C16 hereafter) and the South Pole Telescope \citep[SPT;][]{spaceketal16}. \citet{spaceketal16} co-added the SPT \citep[e.g.,][]{staniszewskietal09} SZ maps around quiescent elliptical galaxies and showed that the observed mean integrated Compton $Y$ parameter is greater than that is expected from simple models without feedback.

R15 prepared two different stacks of Planck SZ maps and cross-correlated them with the SDSS quasar catalog. The two stacks corresponded to two different redshift intervals ($z > 1.5$ and $z < 1.5$). By assuming the high redshift SZ signal to be dominated by quasar feedback, R15 showed that the low redshift signal is consistent with a combination of quasar feedback signal with the same feedback parameters as obtained from the high redshift stack along with an assumed contribution from the gas in the intra-cluster (ICM) medium of quasar host halos. 

\begin{figure}[t]
\begin{center}
\resizebox{8cm}{!}{\includegraphics[angle=-90]{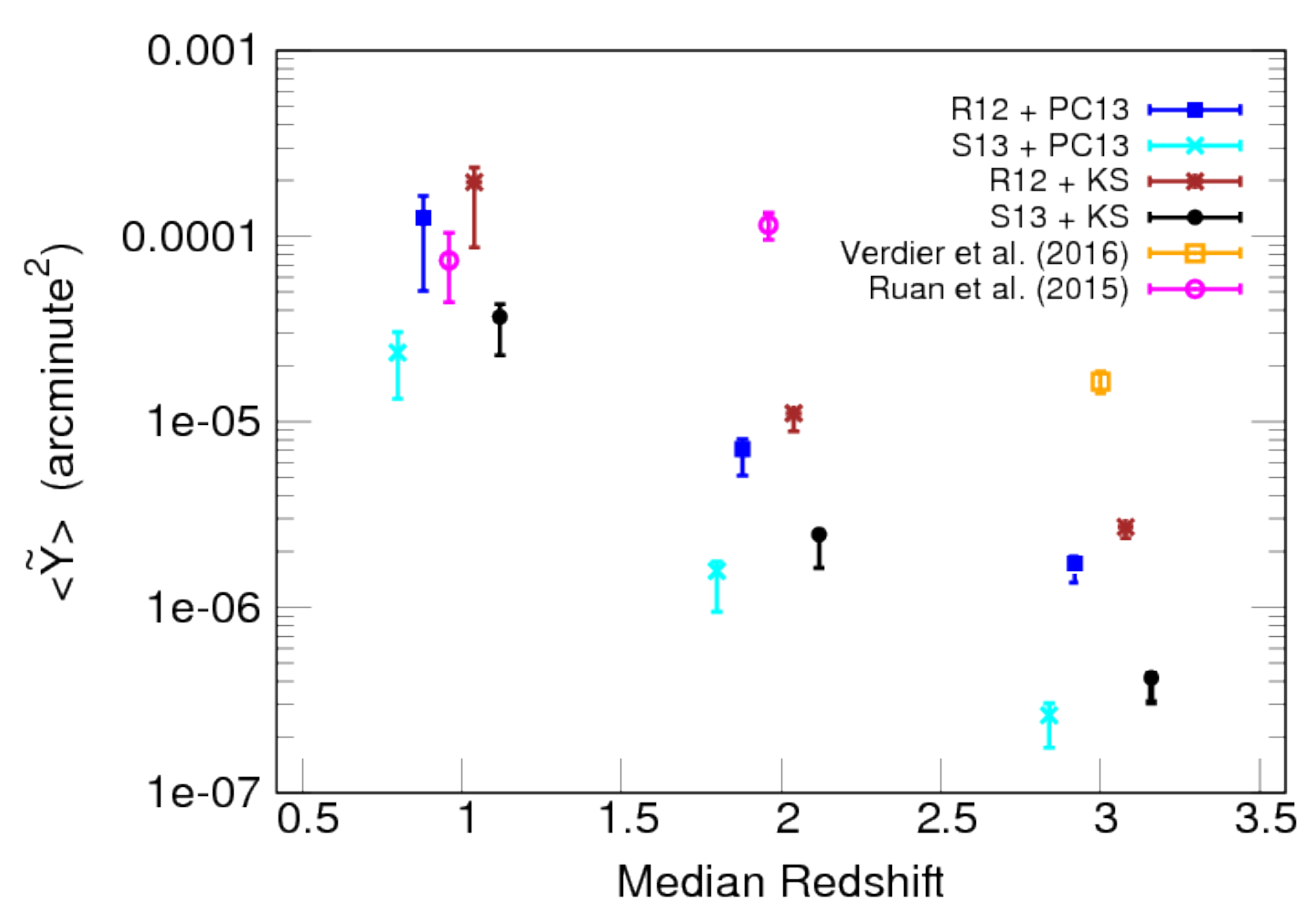}}
\caption{Comparison of R15 (magenta circle) and V16 (orange square) observations for $\langle \tilde{Y} \rangle$  with predictions for the host halo signal considering only gravitational heating of the halo gas from KS and PC13 $Y-M$ relations. Quasars are populated in dark matter halos according to the HOD models of \citet{richardsonetal12} and \citet{shenetal13}. The low redshift signal of R15 is consistent within error limits with the host halo signal predicted using R12 quasar HOD. Using S13 quasar HOD we get a marginally lower value for the host halo signal than what is observed. The high redshift signals of both R15 and V16 are in excess than the host halo contribution predicted from our model.}
\end{center}
\label{fig_1}
\end{figure}

C16 analysed a sample of radio quiet SDSS quasars with redshifts spanning from $0.5 - 3.5$. By stacking Herschel \citep{lutzetal11} and ACT \citep{hincksetal10} data, they derived the spectral energy distribution of quasars in millimeter and far infrared wavebands. Assuming a quasar host halo mass distribution of $(1-5)\times\ 10^{12}\ h^{-1} M_{\odot}$ C16 find that after correcting for dust emission, the observed signal is in excess over that expected from gravitational heating of virialized halo gas alone, favoring a quasar SZ contribution. 

V16 analysed Planck SZ maps cross correlated with SDSS quasars in the redshift range $0.1<z<5$. For their radio quiet quasar sample they detected thermal SZ emission at $2.5<z<4$ with no significant SZ emission at lower redshifts. This is in contradiction to the findings of R15 who detect SZ emission in their $z<1.5$ stack. V16 note that the enhanced signal of R15 is possibly due to improper dust correction. For $z>2.5$, V16 find that the observed signal may or may not be explainable with gravitational heating alone depending on uncertainties in the systematics of the Y-M calibration.

In this paper, we theoretically examine the validity of the detection of SZ effect from quasar feedback in R15, C16 and V16 using a Halo Occupation Distribution (HOD) based approach. We propose that the interpretation of the observed signals is largely dependent on the precise characterization of the mass distribution of quasar hosts, with possible biases arising from poor constraints available on the high mass HOD tail and the largely unexplored redshift dependence of the HOD. For example, using a HOD model of quasars, \citet{c&s15b} show that the signal in R15 can be explained by SZ effect arising from the virialized halo gas alone. We apply a different HOD model proposed by \citealt{chatterjeeetal12} (C12 hereafter) to characterize the SZ contribution from the virialized gas in the quasar host halos and use it to interpret the findings of the cross-correlation signals. 

Our paper is organized as follows. In $\S2$ we briefly describe our model for quantifying the host halo SZ signal. The results of our analysis are presented in $\S3$. We discuss the implications of our results and summarize the main conclusions in $\S4$. Throughout the paper we assume a spatially flat, $\Lambda$CDM cosmology \citep{planckcollaboration15XIII}: $\Omega_{m}=0.31$, $\Omega_{\Lambda}=0.69$, $\Omega_{b}=0.048$, $n_{s}=0.97$, $\sigma_{8}=0.82$, and $h=0.68$.

\section{Theoretical Model}

The thermal SZ temperature distortion at a frequency $\nu$ is given by \citep{s&z70,s&z72}
\begin{equation}
\frac{\Delta T}{T_{0}} = \left[x\coth(x/2)-4\right]y,
\end{equation}
where  $T_{0}$ is the mean CMB temperature (2.73 K) and $x=h\nu/T_0$. The temperature distortion is parametrized by the dimensionless Compton $y$-parameter which is proportional to the line-of-sight integral of the electron pressure ($P_{e}$). The total SZ distortion ($Y$) from a halo can be obtained by integrating the line of sight signal over the solid angle subtended by the halo at the position of the observer. 

The thermal energy of the electron gas can be calculated as
\begin{equation}
E_e=\frac{3}{2} \int^{R_{\rm vir}}_0 P_e\ dV
\end{equation} 
For protons and electrons in thermal equilibrium, the total energy of the ICM gas is
given by
\begin{equation}
E_{\rm tot}= \left(1+\frac{1}{\mu_e}\right)E_e,
\label{E}
\end{equation}
where $\mu_{e}$ is the mean particle weight per electron which we assume as 1.17. To estimate $Y$ and $E_{\rm tot}$ from the virialized gas of quasar host halos in absence of feedback, we use two different methods, namely the analytic prescription of \citet{k&s01} (KS hereafter) and the observed $Y_{500}-M_{500}$ relation of \citealt{planckcollaborationetal13XI} (PC13 hereafter).

The main ingredients of the KS model are as follows: The dark matter density in a halo is described by the self similar NFW profile \citep{nfw95}. The ICM gas is assumed to be in hydrostatic equilibrium, with gas pressure gradient balancing the gravitational attraction of the dark matter. In addition, the gas is assumed to have a polytropic equation of state and gas density is assumed to follow dark matter at halo outskirts. The model does not take gas cooling and star formation into account. For further details on the model we refer the reader to KS.

\begin{figure}[t]
\begin{center}
\resizebox{8cm}{!}{\includegraphics[angle=-90]{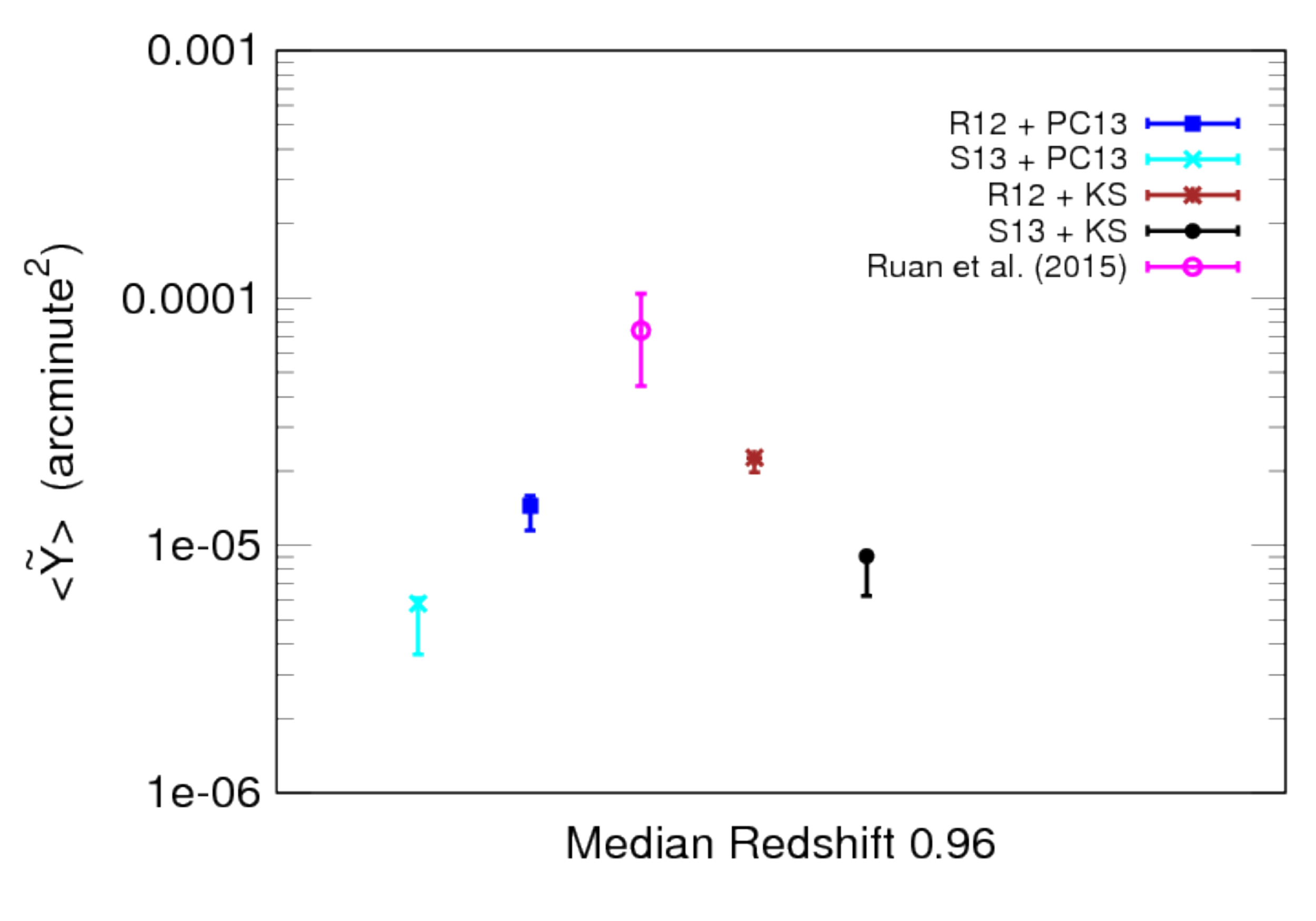}}
\caption{Comparison of R15 (magenta circle) observations for $\langle \tilde{Y} \rangle$ at median redshift $z=0.96$ with predictions for the host halo signal considering only gravitational heating of the virialized gas from KS and PC13 $Y-M$ relations. Quasars are populated in dark matter halos according to the HOD models of \citet{richardsonetal12} and \citet{shenetal13} but the mass of quasar hosts is restricted below $M_{200}=10^{14}\ h^{-1}\ M_{\odot}$. The estimated signals are now lower in magnitude than the observed signal at low redshift indicating that host halo signal modeling at low redshift is sensitive to the high mass HOD tail.}
\end{center}
\label{fig_2}
\end{figure}

PC13 measured the SZ signal from host halos of locally brightest galaxies in the SDSS DR7 catalog as a function of their stellar masses. By assigning halo masses to these galaxies according to the semi-analytic galaxy formation simulation of \citet{guoetal11}, they found a simple power law scaling between $Y_{500}$ and $M_{500}$ extending over a wide range of halo masses from rich clusters down to atleast $M_{500} = 1.34 \times 10^{13} h^{-1} M_{\odot}$. The observed relation after rescaling $Y_{500}$ from each source to a common angular diameter distance of 500\ Mpc is given by
\begin{equation}
\tilde{Y}_{500} =  Y_M \left( \frac{M_{500}}{3.0 \times 10^{14} M_{\odot}}\right)^{\alpha}
\end{equation}
The power law exponent $\alpha$ is fixed at $5/3$ which arises from assuming the ICM gas to be in virial equilibrium with an effective average temperature. The best fit value for the normalization factor $Y_{M}$ is found to be $(0.73 \pm 0.07) \times 10^{-3}\ \rm arcmin^{2}$. We assume the validity of this relation for halo masses below  $M_{500} = 1.34 \times 10^{13} h^{-1} M_{\odot}$ as well. For further details on this model, we refer the reader to PC13.

R15 integrate the Compton-$y$ from each source over a projected radius of $3.75\ h^{-1} Mpc$. On the other hand, PC13 measurements constrain $Y$ in a spherical volume of radius $R_{500}$. Given these differences, R15 note that their observed $\tilde{Y}=1.52\ \tilde{Y}_{500}$. Similarly C16 note that given the ACT beam size, their observed $\tilde{Y}=1.8\ \tilde{Y}_{500}$.

\begin{figure}[t]
\begin{center}
\resizebox{8cm}{!}{\includegraphics[angle=-90]{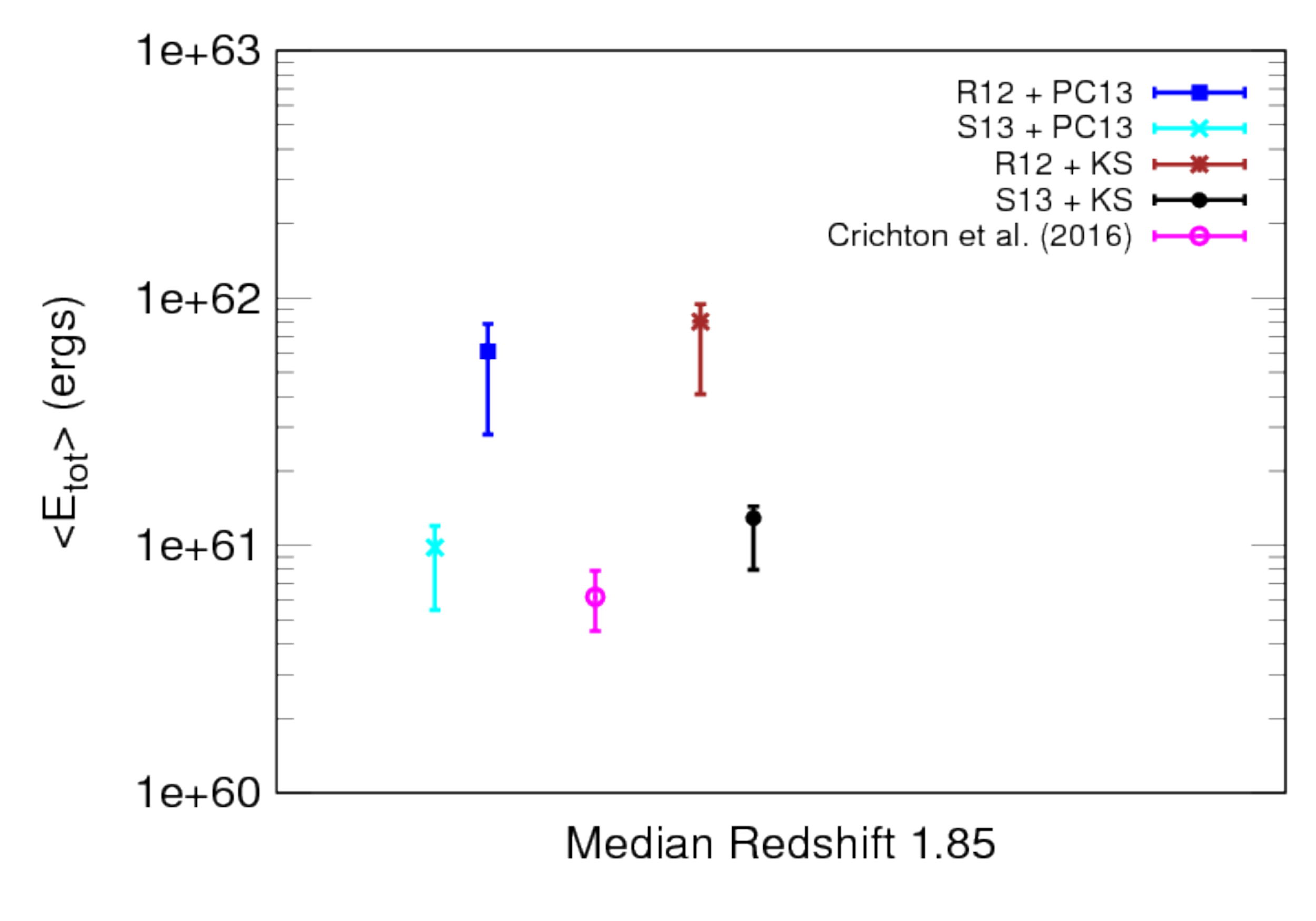}}
\caption{Comparison of C16 observations (magenta circle) for the mean total thermal energy with estimates for the mean total thermal energy of the host halo gas due to gravitational heating alone from KS and PC13 Y-M relations. Quasars are populated in dark matter halos according to the quasar HOD models of R12 and S13. While the host halo signal predicted using S13 HOD is consistent with the observations, that predicted using R12 HOD is above the observed value.}
\end{center}
\label{fig_3}
\end{figure}

To model the average $Y$ in a stack of SZ maps cross-correlated with several quasars, we need to model the host halo distribution of SDSS quasars. To model the distribution of quasar numbers in dark matter halos we follow the HOD formalism \citep[e.g.,][]{b&w02, zhengetal05, zhengetal07, chatterjeeetal12}. The HOD defines the conditional probability that a halo of mass $M$ at redshift $z$ hosts $N$ quasars. We use the 5-parameter HOD model proposed by C12 which is motivated from a study of low luminosity AGN in cosmological hydrodynamic simulations. This particular HOD model has been successfully used to describe the clustering properties of $z=1$ quasars (\citealt{richardsonetal12}, R12 hereafter) and X-ray bright AGN \citep{richardsonetal13} validating the general universality of AGN properties. 

According to this model quasars are divided into two classes, namely central and satellite. The mean occupation function for central quasars is given by a softened step function and that for the satellite ones is a rolling-off power law. The C12 HOD with five free parameters ($M_{\rm min}$, $\sigma_{{\rm log}\ M}$, $M_{1}$, $\alpha$, $M_{\rm cut}$) is given as 
\begin{eqnarray}
\langle N(M)\rangle &=& \frac{1}{2}\left[1+{\rm erf}\left(\frac{{\rm log} M-{\rm log} M_{\rm{min}}}{\sigma_{\rm{log M}}}\right)\right] \nonumber \\ & + & \left(\frac{M}{M_{1}}\right)^{\alpha} \exp \left(-\, \frac{M_{\mathrm{cut}}}{M} \right).
\end{eqnarray}

The average $\tilde{Y}$ (integrated over a range of halo masses and redshifts) is then given as 
\begin{equation}
\langle \tilde{Y} \rangle = \frac{\int_{z_{\rm min}}^{z_{\rm max}}\ \int_{M_{\rm min}}^{M_{\rm max}} \tilde{Y}(M,z)\ \langle N(M) \rangle \frac{dn}{dM}\ dM\ \frac{dV}{dz}\ dz}{\int_{z_{\rm min}}^{z_{\rm max}}\ \int_{M_{\rm min}}^{M_{\rm max}} \ \langle N(M) \rangle \frac{dn}{dM}\ dM\ \frac{dV}{dz}\ dz}, 
\end{equation}
where $dV$ is the comoving volume between redshift $z$ to $z+dz$ and $\frac{dn}{dM}$ is the comoving density of dark matter halos per unit halo mass for which we adopt the \citet{s&t99} mass function. Similarly, the average host halo thermal energy $\langle E_{\rm tot} \rangle$ can be calculated.

We adopt the best-fit HOD parameters from R12 which were obtained by fitting the 2-point correlation function of quasars in the SDSS DR7 catalog at $0.4<z<2.5$ with a median redshift of $1.4$. The best fit values are  $\log M_{\rm min}(h^{-1}M_{\odot}) = 16.46^ {+0.5}_{-0.02}\ $, $\sigma_{\rm logM} = 1.67^{+0.2}_{-0.01}$, $\log M_1 (h^{-1} M_{\odot}) = 12.47^{+3.3}_{-0.6}\ $, $\log M_{\rm cut} (h^{-1}M_{\odot}) = 15.29^{+0.2}_{-0.6} $ and $\alpha = 0.616^{+1.9}_{-0.12}$. R12 interpret their derived HOD to be representative of the ‘true HOD’ of quasars at the median redshift of their clustering sample. However, in our model for the average host halo signal we assume the HOD to be redshift independent and use the R12 parameters for all redshifts in the range $0.1<z<4$. This can possibly lead to bias in our calculations and we discuss this issue in $\S 4$.

We also calculate the average host halo signal at all redshifts using the best-fit parameters from \citet{shenetal13} (S13 hereafter). S13 used the C12 HOD model for fitting the 2-point cross-correlation function of quasars with galaxies in SDSS DR7 catalog at $0.3 < z < 0.9$ with a median redshift of $0.5$ and obtained the best-fit parameters ($M_{\rm min}$, $\sigma_{{\rm log}\ M}$, $M^{'}_{1}$, $\alpha$, $M_{0}$)  as $\log M_{\rm min}(h^{-1}M_{\odot}) = 19.46^ {+0.61}_{-0.64}\ $, $\sigma_{\rm logM} = 2.73^{+2.0}_{-2.1}$, $\log M^{'}_{1} (h^{-1} M_{\odot}) = 16.24^{+0.81}_{-0.51}\ $, $\log M_{0} (h^{-1}M_{\odot}) = 12.74^{+0.86}_{-1.05} $ and $\alpha = 1.19^{+0.37}_{-0.33}$. We however note that due to the narrow redshift range of the S13 sample, R12 HOD will do a comparatively better job in predicting the host halo signal at high redshifts ( $z>1$) than S13 HOD.

\section{Results}
 
In Fig. 1\ we present our results for $\langle \tilde{Y} \rangle$ calculated using both PC13 and KS models corresponding to R15 and V16 quasar samples. For comparison with the low redshift stack of R15, the limits for the redshift integral in Eq.\ 6 are taken as $z_{\rm min}=0.1$, $z_{\rm max}=1.5$. Similarly, for comparison with the high redshift stacks of R15 and V16, the limits are taken as $z_{\rm min}=1.5$, $z_{\rm max}=3.0$ and $z_{\rm min}=2.5$, $z_{\rm max}=4.0$ respectively. For PC13 model, the error bars in the calculated signals represent a combination of the uncertainties in the measured $Y_{500}-M_{500}$ relation and HOD parameters. For KS model, they represent uncertainties in the HOD parameters only. We find that in case of R15 measurements (magenta circle), observed $\langle \tilde{Y} \rangle$ at low redshift ($z<1.5$) is consistent with the signal we can expect from the virialized gas in quasar host halos using both PC13 (blue square) and KS (brown asterisk) models with the R12 host halo distribution ($M_{200}$ varies from $10^{12}\ h^{-1}\ M_{\odot}$ to $10^{15}\ h^{-1}\ M_{\odot}$). Although the best-fit estimates are higher than the observed signal, they overlap within the $1\ \sigma$ error limits.

For the high redshift stack of R15 ($z>1.5$) however, these estimates are about an order of magnitude lower than the observed signal. The absence of massive halos at high redshift together with a redshift independent HOD gives rise to a decrease in the host halo signal. If the S13 HOD parametrization is used ($M_{200}$ varying from $10^{11}\ h^{-1}\ M_{\odot}$ to $10^{15}\ h^{-1}\ M_{\odot}$), we find that the estimated host halo signal for the low redshift stack of R15 (cyan cross for PC13 and black dot for KS) is marginally less than the observed signal. For the high redshift stack these estimates are roughly about 2 orders of magnitude lower than the observed signal.

In the R15 study, all pixels within the angular extent of known clusters are masked. The tSZ maps used in this analysis are derived from \citet{planckcollaborationetal11VIII} early SZ cluster catalog. The least massive clusters in this catalog have masses of the order of $M_{500}=0.6 \times 10^{13} h^{-1} M_{\odot}$ which translates to $M_{200} \sim 10^{14} h^{-1} M_{\odot}$ using $M_{200}$/$M_{500}$=1.6 \citep{duffyetal08}. In Fig.\ 2, we show how applying this mass-cut affects the host halo signal modeling for the low redshift stack of R15. Using R12 HOD with host halo masses varying from $M_{200}=10^{12} h^{-1} M_{\odot}$ to $M_{200}=10^{14} h^{-1} M_{\odot}$, we find that the observed signal at low redshift is at least 3 times higher than the estimate using PC13 model and at least 2 times higher than the estimate using KS model. Using the S13 HOD with host halo masses varying from $10^{11}\ h^{-1}\ M_{\odot}$ to $10^{14}\ h^{-1}\ M_{\odot}$ leads to an even lower prediction for the host halo signal. At higher redshifts on the other hand excluding/including the high mass HOD tail makes almost no statistical difference in the signal modeling. This is expected because of the small number of massive ($M_{200} > 10^{14}\ h^{-1}\ M_{\odot}$) halos at high redshift.

In their study, R15 neglect the contribution from host halo gas in their high redshift sample, assuming it to be entirely dominated by feedback. For their low redshift stack, R15 estimate the contribution from host halo virialized gas by finding the mean $\tilde{Y}$ at $z=0.5$ using the \citet{shenetal13} 5 and 6 parameter HODs and rescale it to $z=0.96$. They find it to be around 2 to 3 times smaller with the 5 and 6 parameter models respectively than the observed signal. From our analysis, we would like to note that the interpretation of the observed signal at low redshift depends on the HOD model used and also on the range of host halo masses considered which would be dependent on the quasar selection function. For the high redshift stack of R15 on the other-hand, there is clear evidence of enhanced signal. Our maximum host halo signal estimate using PC13/KS model with R12 HOD is about an order of magnitude lower than the observed signal. This leaves room for the excess signal being attributed to heating of the ICM by quasar feedback. R15 also draw a similar conclusion for their high redshift stack, however they do not calculate the level of host halo signal that could be present and assume it to be entirely feedback dominated.

Contrary to the R15 results, V16 report that they find no evidence for tSZ signal from quasars below $z<2.5$. They argue that the enhanced signal detected by R15 is due to improper dust subtraction. However, above $z=2.5$, V16 do detect tSZ signal from quasars marked by the orange square in Fig.\ 1. This is obtained by stacking Planck SZ maps cross-correlated with SDSS quasars in the redshift range $2.5<z<4.0$. Planck measures $Y_{500}$, which has been converted to $\tilde{Y}$ using the conversion factor from R15 for the sake of comparison. V16 find the host halo contribution for this sample of quasars by first  obtaining a lower mass bound such that the total number of quasars at $2.5<z<4.0$ is equal to the total number of halos in this redshift range predicted from \citet{tinkeretal08} mass function. They then obtain the mean mass of quasar hosts by integrating the mass function above this mass limit for this redshift range and find it to be $2.18 \times 10^{13}\ h^{-1}\ M_{\odot}$.  This is then used to predict the average host halo signal using the $Y_{500}-M_{500}$ relation from \citealt{arnaudetal10}. 

The \citet{arnaudetal10} relation is calibrated from XMM observations of galaxy clusters and therefore has a mass bias parameter, b which accounts for any bias between the estimated mass and the true halo mass due to deviation from hydrostatic equilibrium. Planck cluster counts suggest a high value of b and taking $b=0.4$, V16 find the that host halo contribution accounts for about $40 \%$ of the observed signal. However, V16 implicitly assume $\langle N(M) \rangle = 1.0$ for all masses above the said lower mass bound. Having total number of quasars equal to total number of halos over a certain redshift range does not necessarily mean that they are equal for each mass and redshift bin. As such it is possible that they could be over estimating the mean mass and thus $\langle Y \rangle$. If b is taken to be zero, then their estimate for the host halo signal is just $1 \sigma$ below the observed signal.

In our model, maximum host halo signal is obtained from using the R12 HOD. With PC13 model for the ICM, we estimate the host halo signal to be at most $\sim 11\ \%$ of the observed signal. Using KS model the estimate is at most $\sim 16\ \%$.  While we find much more excess signal than what can be expected from gravitational heating of gas alone, our results should also be interpreted with caution because of the extrapolation of R12 HOD parameters to $2.5<z<4$. R12 did do a cross-correlation study for $z \sim 3.2$ quasars and found the mean mass of halos hosting central quasars to be $14.1^{+5.8}_{-6.9}\ \times 10^{12}\ h^{-1}\ M_{\odot}$, suggesting an increase in mean mass with redshift. However their results are tentative due to poor statistics. We note that there is a tension between the results of V16 and R12 at low redshifts. We suggest that the poorly constrained contribution of the high mass HOD tail in the clustering studies of R12 and S13 and also the assumption of a redshift independent HOD act as caveats in our theoretical modeling, which in turn remains inconclusive about the true amplitude of the halo gas signal at low redshifts.

In Fig.\ 3, we compare $\langle E_{\rm tot} \rangle$ obtained by C16 (pink circle) with the theoretical estimates from KS and PC13 models using R12 and S13 HODs. The limits for the redshift integral in Eq.\ 6 are taken as $z_{\rm min}=0.5$ and $z_{\rm max}=3.5$. We find that using the R12 host halo distribution , our estimate for the host halo signal (blue square for PC13 and brown asterisk for KS) is greater than the observed signal. Using the S13 HOD (cyan cross for PC13 and black dot for KS), our estimate is consistent with the observed signal within error-limits, leaving no room for excess heating due to feedback. By assuming host halo masses of quasars to vary to $(1-5) \times 10^{12}\ h^{-1}\ M_{\odot}$, C16 noted that gravitational heating could account for only $30\%$ of the observed signal. However, modeling the average host halo signal using the R12 HOD at the median redshift of their sample ($z=1.85$), they found that the observed signal could be explained with gravitational heating alone.

From Figs. 1,\ 2\ and 3\ we note that the KS model systematically over-estimates the host halo signal in all cases because of the absence of gas cooling and star formation. Thus, the level of host halo signal weakly depends on the details of the Y-M calibration with a more stronger HOD dependence as in all cases S13 HOD gives a much lower estimate.

\section{Discussion}
Several observational evidences and theoretical models suggest that there exists a strong link between galaxy evolution and the growth of super massive black holes (SMBH) at galaxy centers \citep[e.g.,][]{m&f01, tremaineetal02, grahametal11}. The key ingredient to this link has been attributed to feedback from the central black hole \citep[e.g.,][]{shankaretal04,hopkinsetal06,lapietal06,hopkinsetal08, dimatteoetal08, b&s09,c&o07}. Different observational probes, such as the SZ effect, have been proposed in the literature to systematically study the effect of feedback on the gas in the ICM \citep[see references in][]{chatterjeeetal15b}. The noise sensitivity and the angular resolution of current and proposed CMB experiments are just in the limit of detecting the SZ signal from quasar feedback.

However, a key issue related to this detection involves disentangling the signal from the SZ effect arising from the virialized gas in the host dark matter halos of AGN. In recent work the effect has been mostly studied with bright quasars due to their abundance in large surveys. We observe that while estimating the SZ contribution from quasar hosts, most studies assume the host halos of quasars to lie in the range of $\approx 10^{12}M_{\odot}$, which happens to be the peak in the mass distribution of quasar hosts (see R12). In this study, we emphasize the importance of understanding the implication of the full host halo mass distribution of quasars in characterizing the SZ signal from the virialized gas. Recent measurements of quasar HOD allows us to account for this correction. 

An important component to this approach lies in modeling the redshift dependence of the quasar HOD which is yet to be understood in numerical simulations (or semi-analytic work). For low-luminosity AGN, C12 (HOD model used in this work) report a redshift dependence of the HOD parameters but the results are tentative due to lack of statistics. We also note that while fitting the 2PCF of $z \approx 1$ quasars, R12 interpret the derived HOD to be representative of the `true HOD' of quasars at the median redshift of their clustering sample. R12 validates this assumption by claiming that the true HOD at the median redshift will be represented by the measured HOD from the 2PCF constructed over a wide redshift range, if the redshift evolution of the 2PCF (and the HOD thereof) is confined within the statistical errors of the 2PCF. In our analysis we adopted a redshift independent HOD following the argument in R12. But, we note that once the statistical power of the 2PCF measurements increase substantially, the redshift evolution of the HOD needs to be accounted in studies involving quasar hosts. We also note that there can be possible theoretical bias regarding the model of the HOD itself, since the current 2PCF measurements are degenerate to varying theoretical models.  We refer the reader to R12 for more discussion on this. Direct measurement techniques of the quasar HOD \citep{chatterjeeetal13} can provide alternative ways to break this degeneracy.

Besides the HOD description, another major component for disentangling the two signals is governed by the assumptions on feedback models. Most studies assume a thermal mode of feedback and predict that the hot gas coming from the central SMBH is responsible for increasing the temperature of the plasma while decreasing the gas density around it \citep[e.g.,][]{chatterjeeetal08, scannapiecoetal08, battagliaetal10}. These studies show that the result of these two opposite effects is still responsible in enhancing the SZ signal from what is expected due to virialized gas. As discussed before, these predictions are model dependent and the possibility of a decrement of the signal is yet to be ruled out. We note that within the restrictions of the assumed model of feedback \citet{chatterjeeetal08} predicts that the SZ effect from quasars is more enhanced at high redshift when the black hole is more active. We do observe a similar trend in the results of R15 where they claim that the enhanced SZ signal from the high redshift stack is a result of quasar feedback. 

Another important issue is related with the gas model we use to obtain the SZ signal from quasar hosts. The K-S model assumes a polytropic equation of state and neglects the intricate astrophysics in galaxy groups and clusters. Improvement over the K-S model has been proposed by different authors \citep[e.g.,][]{bodeetal09, shawetal10} and for more accurate modeling of the ICM we would need to consider these upgraded models. But we note that the observationally derived PC13 scaling provides similar results. We hence propose that the constraints on the quasar HOD seem to be the dominating source of uncertainty in determining the host halo SZ signal and hence the exclusion of more detailed cluster physics does not alter our conclusion.  

Our study thus shows that within the framework of our simplified assumptions it is difficult to disentangle the halo signal from the feedback signal at low redshifts. However at higher redshifts ($z>2.5$), there is evidence for enhanced signal that cannot be explained with gravitational heating alone. Future AGN surveys will enhance the statistical power of 2PCF measurements providing tighter constraints on quasar/AGN HOD. That would allow us to estimate the SZ signal from quasar hosts with better precision. The pre-requisite for interpreting those results links to a superior theoretical understanding of the HOD itself (e.g., accurate redshift evolution). We thus emphasize the need for both theoretical and observational studies of the HOD for better understanding of the relationship between quasars and their host dark matter halos. Our analysis on the need for precise determination of the HOD is not only important for determination of quasar SZ signal but is essential for using quasars as probes of the high redshift Universe.

\section*{Acknowledgments}
The authors thank the anonymous referee for many useful comments that helped in improvement of the paper and Yue Shen for sharing some of his data products that helped in the HOD comparison. DDC thanks Kanan Kumar Datta for his support, Saumyadip Samui for his help with some aspects of the analysis and acknowledges the Department of Science and Technology, Govt. of India for financial support through the INSPIRE scholarship. DDC also acknowledges useful discussions with Daisuke Nagai and Kaustav Mitra that helped in interpreting some results in the paper. SC acknowledges partial support from the University Grants Commission through the start-up grant and Presidency University Kolkata through the FRPDF grant. SC is grateful to the Inter University Center for Astronomy and Astrophysics (IUCAA) for providing infra-structural and financial support along with local hospitality through the IUCAA-associateship program. 

\bibliography{ms}{}
\end{document}